\documentclass{elsart}

\usepackage{natbib}

\usepackage{graphicx}
\usepackage{amssymb}

\begin{document}

\begin{frontmatter}

% Title, authors and addresses

% use the thanksref command within \title, \author or \address for footnotes;
% use the corauthref command within \author for corresponding author footnotes;
% use the ead command for the email address,
% and the form \ead[url] for the home page:
% \title{Title\thanksref{label1}}
% \thanks[label1]{}
% \author{Name\corauthref{cor1}\thanksref{label2}}
% \ead{email address}
% \ead[url]{home page}
% \thanks[label2]{}
% \corauth[cor1]{}
% \address{Address\thanksref{label3}}
% \thanks[label3]{}

\title{X-ray Variability and Emission Process of the Radio Jet in M87}

% use optional labels to link authors explicitly to addresses:
% \author[label1,label2]{}
% \address[label1]{}
% \address[label2]{}

\author{ D. E. Harris}

\address{SAO, 60 Garden St., Cambridge, MA 02138 USA}
\ead{harris@cfa.harvard.edu}

\begin{abstract}

We monitored the M87 jet with the ACIS-S detector on Chandra with 5
observations between 2002 Jan and 2002 Jul.  Our goal was to determine
the presence and degree of variability in morphology, intensity, and
spectral parameters.  We find strong variability of the core and
HST-1, the knot lying 0.8$^{\prime\prime}$ from the core.  These
observations were designed to constrain the X-ray emission process:
whereas synchrotron emission would necessitate the presence of
extremely high energy electrons with a halflife of a few years or
less, inverse Compton emission from a relativistic jet would arise
from low energy electrons with very long halflives.  Currently, all
indications point to a synchrotron process for the X-ray emission from
the M87 jet.  We give key parameters for a ``modest beaming''
synchrotron model.

\end{abstract}

\begin{keyword}
% keywords here, in the form: keyword \sep keyword
M87 \sep relativistic jets \sep X-ray variability

% PACS codes here, in the form: \PACS code \sep code

\end{keyword}

\end{frontmatter}

% main text

\section{Introduction}

Although there is general agreement that X-ray emission from radio
jets arises from non-thermal processes, both synchrotron emission and
inverse Compton (IC) emission are indicated for different, and
occasionally the same, radio features (Celotti et al. 2001, Marshall et
al. 2001, Sambruna et al. 2001).  Here we present strong evidence that
synchrotron emission is responsible for the X-ray emission from the
M87 jet and demonstrate a reasonable combination of source parameters
for HST-1, the jet feature lying 0.8$^{\prime\prime}$ from the nucleus.

We take the distance to M87 to be 16 Mpc so that one arcsec~=~77pc.
We use the normal convention for spectral index: flux density,
S~$\propto~\nu^{-\alpha}$.

\section{Data Processing and Results}

We used a 1/8th subarray on the back illuminated ACIS-S3 chip with
0.4s readout to mitigate pileup; the same configuration used for a
12ks archival observation of 2000Jul.  A 'standard' reduction system,
based on the CIAO threads, was developed to ensure that each dataset
was processed identically.  We believe that the key to reliable
photometry for closely spaced features is to remove the pixel
randomization and re-grid to 1/10th pixel.  This process recovers the
true resolution of the Chandra Observatory.  The event files were then
binned and filtered into three energy bands: soft = 0.2-0.75keV;
medium = 0.75-2 keV; and hard = 2-6 keV.  The bands were chosen to
minimize the change in effective area across each band, although this
was not possible for the soft band.  We then created corresponding
exposure maps and divided the images by these to obtain flux maps.  By
testing on a fixed area of hot cluster gas, we recovered the quantum
efficiency (QE) degradation of the ACIS CCD.  Multiplying each band
map by h$\nu$ of the mean energy plus the correction factor for QE, we
obtained flux maps comparable to the 2000 observation.

\begin{figure}
%  \rotatebox{90}{
%  \resizebox{8.0cm}{!}
\includegraphics[scale=0.7]{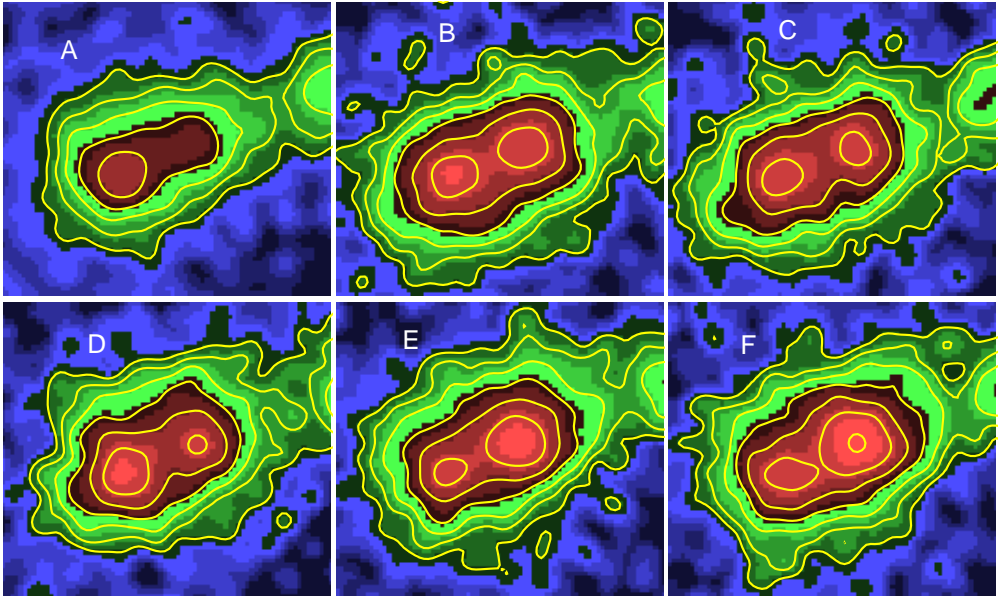} \caption{The core and HST-1: (A)
        is the archival observation of 2000Jul and our 5 monitoring
        observations of 2002 follow.  The contours increase by factors
        of two with the lowest contour level being
        2$\times$10$^{-16}$~erg~cm$^{-2}$~s$^{-1}$~per pixel in the
        0.2 to 6 keV band.  The data were regridded to a pixel size of
        0.0492$^{\prime\prime}$; and a Gaussian smoothing function of
        FWHM=0.25$^{\prime\prime}$ was applied.}
\label{fig:6pan}
\end{figure}

\begin{figure}
%  \rotatebox{90}{
%  \resizebox{8.0cm}{!}{
\includegraphics[scale=0.5, angle=-90]{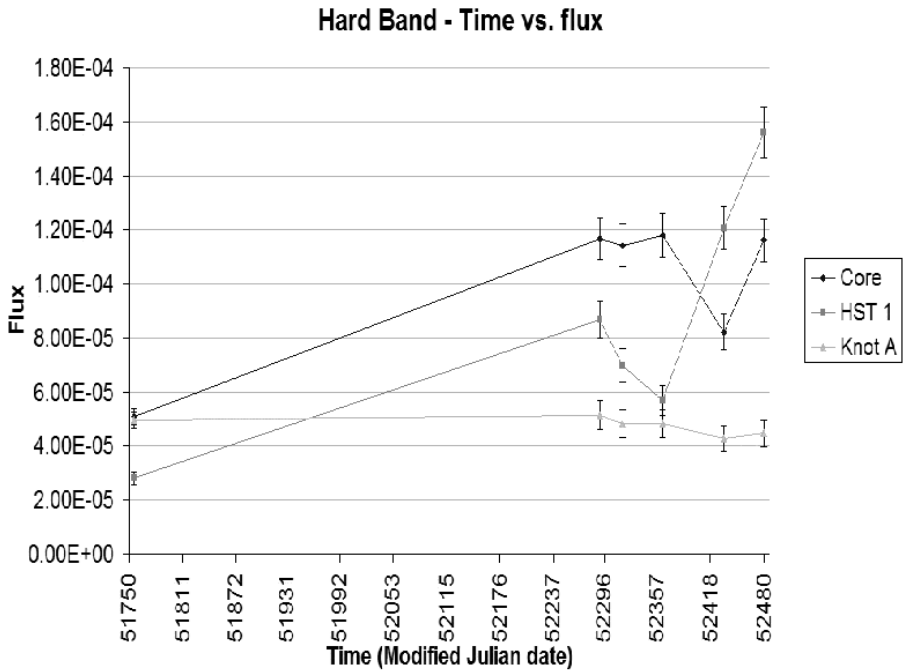}

\caption{Hard band lightcurves for the M87 core, HST-1, and knot A.  The
observation at the left is from 2000Jul; those to the right from the
2002 season.  Error bars are SQRT(N) type, based on total counts in
the measuring apertures prior to background subtraction.  The energy
band is 2 to 6 keV. \label{fig:photom}}

% \caption{The hard band light curves for the core, HST-1, and knot A.}
\label{fg:lc}
\end{figure}

The basic results for the core and HST-1 are shown in
fig.~\ref{fig:6pan} and strong intensity variability can be seen.
Using small circular apertures for the core, knot A, and HST-1, we
obtained the relative fluxes shown in figure~\ref{fig:photom}.  HST-1
increased by a factor of two in the last 116 days covered by our
observations.  The light travel time for this interval is 0.1pc which
may be compared with our inherent resolution of order
0.3$^{\prime\prime}$ or 23pc at the distance of M87.

% \section{Measurements}

\section{HST-1}

We take our variability data of HST-1 to support the synchrotron
emission model since when the flux is decreasing, there is a larger
percentage drop in the hard band than in the medium or soft bands, and
the flux decrease is consistent with halflives of order a year rather
than a value of order 10,000 years corresponding to the low energy
electrons which would produce inverse Compton (IC) X-rays.

We used a single power law to fit the optical (Perlman et al. 2001)
and X-ray flux densities of HST-1 from $\nu_1~=~10^9$~Hz to
$\nu_2~=~10^{18}$~Hz with $\alpha$~=~0.68.  With this spectrum, we
then derived synchrotron parameters based on the usual equipartition
assumptions (Pacholczyk, 1970) for two alternative source sizes:
r=0.3$^{\prime\prime}$ and a radius corresponding to the light travel
time of 116 days.  For each choice, we examined parameters for beaming
factors, $\delta$ ranging from 1 to 16.  For those parameters
requiring a value of the bulk relativistic velocity, we assumed
$\Gamma=\delta$.

The detailed results of these models will be given in Harris et
al. (2003), but here we summarize the results by noting that for the
large source size of 0.3$^{\prime\prime}$, values of the halflife,
$\tau_{\frac{1}{2}}$, for the highest energy electrons are greater
than a year for all values of $\delta~<$~16.  Since large $\delta$'s
require very small angles of the jet axis to the line of sight
(l.o.s.), the data support the smaller diameter alternatives.

For models with the source size given by the light travel time, small
values of $\delta$ have values of $\tau_{\frac{1}{2}}$ too small to
match our data and large values of $\delta$ predict
$\tau_{\frac{1}{2}}$ significantly larger than a year.  The parameters
that best fit the data are provided by $\delta$~=~4.  For this model,
the characteristic angle to the l.o.s. is 14$^{\circ}$, consistent
with canonical values from radio considerations; a magnetic field
strength of 1~mG; and a total energy requirement (field plus
particles) of 10$^{48}$~ergs.  In the jet frame, the time required to
double the intensity (and thus double the total energy reservoir) is
$\approx~4\times10^7$s, which translates into a very modest 'drain' on
jet power of $<~10^{41}$ erg~s$^{-1}$ The photon energy density in the
jet frame is dominated by the synchrotron spectrum, but is of order
5\% of the magnetic field energy density.

% @@
% \section{Summary}

Since synchrotron self-Compton models under predict the observed X-ray
emission by a large factor and IC/CMB models require unreasonable
values of $\Gamma$ and angles to the l.o.s. (Harris \& Krawczynski,
2002), the ``modest beaming'' synchrotron model appears to be the
model of choice.

% \section{Acknowledgments}

Colleagues for this project are J. Biretta, W. Junor, E. Perlman,
W. Sparks, \& A. Wilson; they will be co-authors of a longer paper
(Harris et al., 2003).  We thank O. Stohlman for assistance in the
data reduction.  This work was supported by NASA contract NAS8-39073
and grant GO2-3144X.

% The Appendices part is started with the command \appendix;
% appendix sections are then done as normal sections
% \appendix

% \section{}
% \label{}

% Bibliographic references with the natbib package:
% Parenthetical: \citep{Bai92} produces (Bailyn 1992).
% Textual: \citet{Bai95} produces Bailyn et al. (1995).
% An affix and part of a reference:
%   \citep[e.g.][Ch. 2]{Bar76}
%   produces (e.g. Barnes et al. 1976, Ch. 2).

\end{document}